\documentclass[conference]{IEEEtran}
\IEEEoverridecommandlockouts
\usepackage{graphics} 
\usepackage{epsfig} 
\usepackage{mathptmx} 
\usepackage{times} 

\usepackage{multirow}
\usepackage{booktabs} 
\usepackage{subcaption}
\usepackage{cite}
\usepackage{amsmath,amssymb,amsfonts}
\usepackage{algorithmic}
\usepackage{graphicx}
\usepackage{textcomp}
\usepackage{xcolor}
\usepackage{float} 

\def\BibTeX{{\rm B\kern-.05em{\sc i\kern-.025em b}\kern-.08em
    T\kern-.1667em\lower.7ex\hbox{E}\kern-.125emX}}
\begin{document}

\title{ARIES-Mission2: A Zero-Shot Vision-Language-Action Framework for Fast Large-Scale Aerial Mission Generation
\thanks{This work was supported by the grant from Macao Polytechnic University (RP/FCA-06/2026) and Macao Science and Technology Development Fund (FDCT-MOST: 0018/2025/AMJ).}
}

\author{\IEEEauthorblockN{1\textsuperscript{st} Junhao Wei}
\IEEEauthorblockA{\textit{Faculty of Applied Sciences} \\
\textit{Macao Polytechnic University}\\
Macao, China \\
p2312195@mpu.edu.mo}
\and
\and
\IEEEauthorblockN{2\textsuperscript{nd} Yanxiao Li}
\IEEEauthorblockA{\textit{Faculty of Applied Sciences} \\
\textit{Macao Polytechnic University}\\
Macao, China \\
P2525981@mpu.edu.mo}
\and
\IEEEauthorblockN{3\textsuperscript{rd} Haochen Li}
\IEEEauthorblockA{\textit{Faculty of Applied Sciences} \\
\textit{Macao Polytechnic University}\\
Macao, China \\
p2523372@mpu.edu.mo}
\and
\IEEEauthorblockN{4\textsuperscript{th} Yifu Zhao}
\IEEEauthorblockA{\textit{Faculty of Applied Sciences} \\
\textit{Macao Polytechnic University}\\
Macao, China \\
p2523269@mpu.edu.mo}
\and
\IEEEauthorblockN{5\textsuperscript{th} Dexing Yao}
\IEEEauthorblockA{\textit{Faculty of Applied Sciences} \\
\textit{Macao Polytechnic University}\\
Macao, China \\
P2522978@mpu.edu.mo}
\and
\IEEEauthorblockN{6\textsuperscript{th} Baili Lu}
\IEEEauthorblockA{\textit{Faculty of Applied Sciences} \\
\textit{Macao Polytechnic University}\\
Macao, China \\
18023304003@163.com}
\and
\IEEEauthorblockN{7\textsuperscript{th} Zikun Li}
\IEEEauthorblockA{\textit{School of Economics and Management} \\
\textit{South China Normal University}\\
Guangzhou, China \\
20190731013@m.scnu.edu.cn}
\and
\IEEEauthorblockN{8\textsuperscript{th} Yapeng Wang}
\IEEEauthorblockA{\textit{Faculty of Applied Sciences} \\
\textit{Macao Polytechnic University}\\
Macao, China \\
yapengwang@mpu.edu.mo}
\and
\IEEEauthorblockN{9\textsuperscript{th} Sio-Kei Im}
\IEEEauthorblockA{\textit{Macao Polytechnic University}\\
Macao, China \\
marcusim@mpu.edu.mo}
\and
\IEEEauthorblockN{10\textsuperscript{th} Dingcheng Yang}
\IEEEauthorblockA{\textit{Information Engineering School} \\
\textit{Nanchang University}\\
Nanchang, China \\
yangdingcheng@ncu.edu.cn}
\and
\IEEEauthorblockN{11\textsuperscript{th} Xu Yang*}
\IEEEauthorblockA{\textit{Faculty of Applied Sciences} \\
\textit{Macao Polytechnic University}\\
Macao, China \\
xuyang@mpu.edu.mo}
*Corresponding author
}

\maketitle

\begin{abstract}
Multimodal Large Language Models (MLLMs) have shown strong semantic understanding capabilities, but their direct use in low-altitude Unmanned Aerial Vehicle (UAV) mission generation remains limited by weak spatial optimization and inefficient route planning. To address this issue, we propose ARIES-Mission2, a zero-shot Vision-Language-Action (VLA) framework that decouples visual-semantic perception from physical route optimization. Given natural-language instructions and satellite imagery, ARIES-Mission2 first uses DeepSeek-V3 for task parsing and Molmo-7B for zero-shot target grounding, and then converts detected pixel locations into GPS waypoints through geospatial interpolation. To reduce the redundant backtracking caused by raw VLM-generated visiting orders, the back end formulates multi-target UAV traversal as a Traveling Salesperson Problem (TSP) and compares four candidate routes, including the raw VLM order and the routes optimized by PSO, GPSO, and IPSO. The minimum-cost closed-loop route is then selected for mission generation. Experiments on the UAV-VLPA-nano-30 benchmark show that ARIES-Mission2 achieves a total flight distance of 62.43 km, reducing the route length by 21.6\% compared with the unoptimized VLA baseline (79.66 km) and by 9.5\% compared with manual human planning (69.00 km). The complete 30-task workflow takes 575.40 s, averaging 19.18 s per task, which is approximately 3.6 times faster than human expert planning. Component-level timing shows that VLM inference dominates the runtime with 19.02 s per task, while the TSP solver requires only 0.16 s per task. Scalability analysis further indicates that the TSP module maintains lower growth in computation time as the number of targets increases. These results demonstrate that coupling zero-shot visual grounding with lightweight route optimization can improve both efficiency and route quality for large-scale UAV mission generation.
\end{abstract}

\begin{IEEEkeywords}
UAV, LLM, VLA, metaheuristics, mission generation
\end{IEEEkeywords}

\section{Introduction}
In recent years, Unmanned Aerial Vehicles (UAVs) have become indispensable in the low-altitude economy and wide-area inspections. To enable natural human-robot interaction, Vision-Language-Action (VLA) models have emerged to map multimodal inputs directly into physical actions. This has further evolved into Vision-Language Navigation (VLN), requiring UAVs to conduct precise spatial reasoning and long-horizon exploration in complex environments.

Academic exploration of Large Language Models (LLMs) in UAV control has been intensive. For instance, TypeFly (2023) addressed LLM generation latency by integrating the MiniSpec language, reducing response times by 62\% \cite{Typefly}. Similarly, CityNavAgent (2025) utilized a Hierarchical Semantic Planning Module (HSPM) to decompose long-horizon tasks \cite{CityNavAgent}. However, the MM-UAVBench benchmark (2025) systematically evaluated 16 models and revealed significant bottlenecks in perception and planning, noting that current MLLMs often struggle to output physically feasible navigation strategies due to spatial biases \cite{MM-UAVBench}.

This 'semantic-to-physical' disconnect highlights a core deficiency in spatial operations research. While VLMs can extract target coordinates via zero-shot capabilities, they lack the geometric optimization needed for efficient visiting sequences, often resulting in redundant backtracking. Fundamentally, multi-target UAV traversal is a Traveling Salesperson Problem (TSP) \cite{landscape}. To compensate for MLLMs' shortcomings in combinatorial optimization, we introduce metaheuristic algorithms, which provide a robust framework for finding near-optimal solutions to NP-Hard problems \cite{GWOA} \cite{GLNWOA} \cite{WASHH} \cite{low} within reasonable computational overhead \cite{IPSO} \cite{LSEWOA} \cite{MRBMO} \cite{ESTGWOA} \cite{DHCRWOA}.

To bridge this loop, this paper proposes ARIES-Mission2, which adopts a decoupled 'front-end visual semantic inference + back-end physical optimization' architecture. The front end utilizes an LLM/VLM pipeline for target extraction, while the back end integrates three metaheuristic algorithms (PSO, GPSO, and IPSO) for route planning. Our main contributions are:

\begin{itemize}
    \item \textbf{Proposed the ARIES-Mission2 framework:} We decouple VLM perception from a metaheuristic TSP optimizer to resolve LLMs' inherent defects in geometric reasoning.
    \item \textbf{Achieved breakthrough flight efficiency:} On the UAV-VLPA-nano-30 benchmark, ARIES-Mission2 reduced total flight distance to 62.43 km, significantly outperforming the pure VLA baseline (79.66 km) and human experts (69 km).
    \item \textbf{Realized highly efficient automated deployment:} The system completed 30 complex scenarios in under 10 minutes (VLM: 9.51 min, TSP: 0.08 min), operating far faster than the 35 minutes required by manual expert planning.
\end{itemize}

\begin{figure*}[t]
    \centering
    \includegraphics[width=\textwidth]{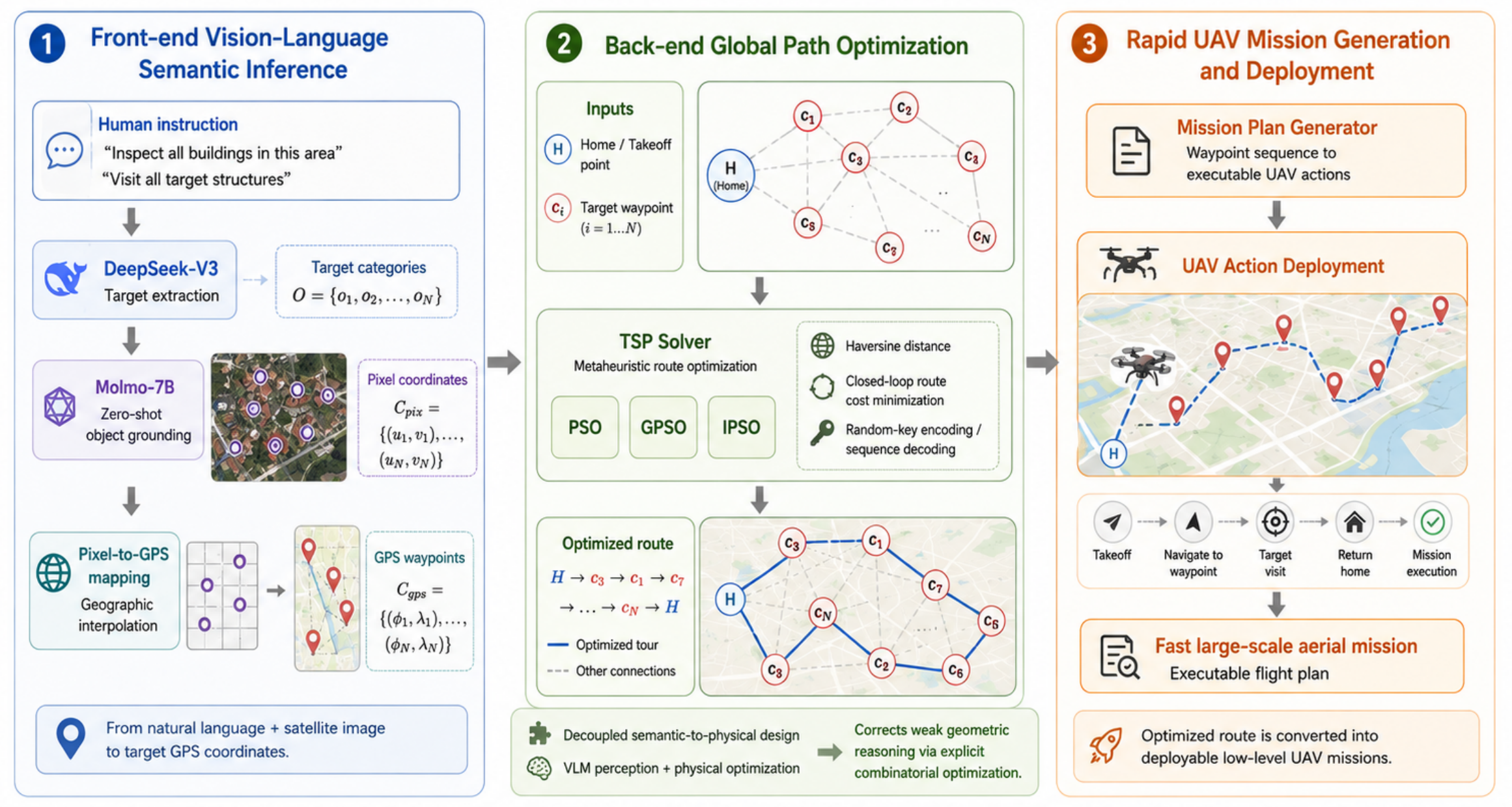}
    \caption{Overview of the proposed ARIES-Mission2 framework. The system decouples zero-shot visual-semantic perception from physical route optimization. Natural-language instructions and satellite imagery are first processed by a large language model and a vision-language grounding model to extract target categories and pixel-level target locations, which are then converted into GPS waypoints. The back-end TSP solver optimizes the visiting order using metaheuristic route search, and the resulting waypoint sequence is transformed into executable UAV mission actions.}
    \label{yuanli}
\end{figure*}

\section{The proposed framework}
To address the shortcomings of large multimodal models in geometric computation for complex long-horizon spatial reasoning, this paper proposes a rapid large-scale UAV mission generation framework (ARIES-Mission2), as shown in Fig.~\ref{yuanli}. This framework adopts a decoupled architecture, dividing the system into three core stages: front-end vision-language inference based on multimodal large models, back-end global path optimization by a TSP Solver based on metaheuristic algorithms, and the rapid automatic generation and deployment of low-level UAV actions.

\subsection{Vision-Language Semantic Inference}
The core task of the front-end module is to transform unstructured natural language instructions input by human operators into discrete target point coordinates with precise geographic information. Given a sequence of natural language instruction sets $\mathcal{U}=\{w_1, w_2, \dots, w_m\}$ input by a human, the system first invokes a large language model (DeepSeek V3) as a semantic parser to extract the task targets:
\begin{equation}
    \mathcal{O} = \Phi_{DeepSeek}(\mathcal{U}) = \{o_1, o_2, \dots, o_N\}
     \label{eq1}
\end{equation}
where $\mathcal{O}$ represents the specific target categories (e.g., 'buildings') extracted from the instructions, and $N$ is the total number of such targets subsequently detected in the image.

Subsequently, the Vision-Language Model (Molmo-7B) receives the extracted targets $\mathcal{O}$ and the corresponding satellite image $\mathcal{I}_{sat}$, precisely locating the pixel coordinate set of all targets on the 2D image plane:
\begin{equation}
    \mathbf{C}^{pix} = \Psi_{Molmo}(\mathcal{O}, \mathcal{I}_{sat}) = \{(u_1, v_1), (u_2, v_2), \dots, (u_N, v_N)\}
     \label{eq2}
\end{equation}
To eliminate physical computation deviations caused by image resolution and scale, the system extracts the geographic metadata (latitude and longitude boundaries) embedded in the satellite image and utilizes a linear interpolation function $\Gamma_{geo}$ to map the 2D pixel coordinates into real-world Global Positioning System (GPS) coordinates:
\begin{equation}
    \mathbf{C}^{gps} = \Gamma_{geo}(\mathbf{C}^{pix}) = \{(\varphi_1, \lambda_1), (\varphi_2, \lambda_2), \dots, (\varphi_N, \lambda_N)\}
     \label{eq3}
\end{equation}
where $\varphi_i$ and $\lambda_i$ represent the true latitude and longitude of target $i$, respectively. This step establishes a highly precise geographic baseline for subsequent physical-level optimal path planning.

\subsection{TSP Solver based on Metaheuristics}
\begin{figure}[htbp]
    \centering
    \includegraphics[width=0.5\textwidth]{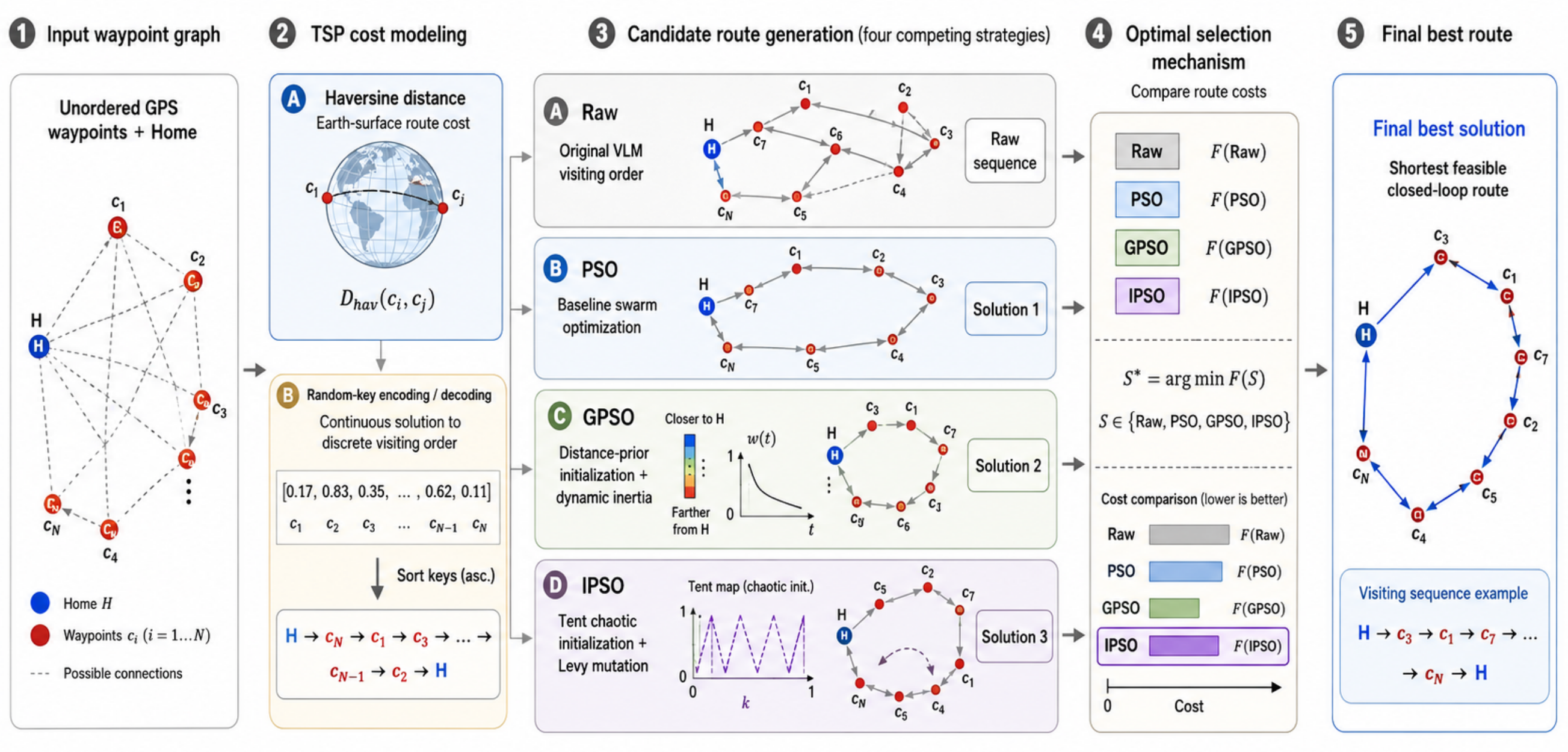}
    \caption{Internal workflow of the TSP solver. Given the UAV home position and unordered GPS waypoints, the solver computes geographic route costs using the Haversine distance and decodes continuous random-key representations into discrete visiting sequences. Four competing candidate routes are evaluated, including the raw VLM visiting order and the optimized routes generated by PSO, GPSO, and IPSO. The final closed-loop route is selected by minimizing the route cost over all candidates.}
    \label{tsp}
\end{figure}

The extracted target physical coordinate set $\mathbf{C}^{gps}$, combined with the UAV takeoff point (Home, denoted as $H(\varphi_0, \lambda_0)$), constitutes a typical Traveling Salesperson Problem (TSP) network. Fig.~\ref{tsp} is a schematic diagram of the \textit{\textbf{TSP Solver}} with metaheuristic algorithms for solving the TSP problem in the ARIES-Mission2 framework.
To calculate the most realistic flight cost, the TSP Solver internally employs the Haversine formula to compute the Earth's spherical physical distance. For any two points $A(\varphi_A, \lambda_A)$ and $B(\varphi_B, \lambda_B)$, their physical distance $D_{hav}(A, B)$ is calculated as follows:
\begin{equation}
    a = \sin^2\left(\frac{\Delta\varphi}{2}\right) + \cos(\varphi_A)\cos(\varphi_B)\sin^2\left(\frac{\Delta\lambda}{2}\right)
     \label{eq4}
\end{equation}

\begin{equation}
    D_{hav}(A, B) = 2R \cdot \text{arctan2}(\sqrt{a}, \sqrt{1-a})
     \label{eq5}
\end{equation}
where $R = 6371.0$ km is the Earth's radius.For any visiting sequence encoding $\mathbf{X} = [x_1, x_2, \dots, x_N]$ containing $N$ targets (decoded into a discrete sequence in ascending order via the Random Key representation), the overall cost function $F(\mathbf{X})$ of the closed-loop flight is defined as:
\begin{equation}
    F(\mathbf{X}) = D_{hav}(H, \mathbf{c}_{x_1}) + \sum_{k=1}^{N-1} D_{hav}(\mathbf{c}_{x_k}, \mathbf{c}_{x_{k+1}}) + D_{hav}(\mathbf{c}_{x_N}, H)
     \label{eq6}
\end{equation}

Since TSP is a classic NP-Hard problem, the search space experiences a factorial explosion as the number of target nodes $N$ increases. Therefore, this study proposes two improved metaheuristic algorithms (GPSO and IPSO). The back-end TSP Solver of this framework integrates these two improved Particle Swarm Optimization algorithms (GPSO and IPSO) along with the basic PSO \cite{PSO} to rapidly compute the global near-optimal flight path.

\subsection{PSO}
Within the integrated algorithmic architecture, each evolution of the particle swarm relies on the core velocity and position update equations. In each iteration, based on its personal historical best position $\mathbf{Pbest}_i$ and the current population's global best position $\mathbf{Gbest}$, the particle calculates a new velocity $\mathbf{V}_{i}^{(t+1)}$ and updates its position $\mathbf{X}_{i}^{(t+1)}$:
\begin{equation}
    \mathbf{V}_{i}^{(t+1)} =  \mathbf{V}_{i}^{(t)} + c_1 r_1 (\mathbf{Pbest}_i - \mathbf{X}_{i}^{(t)}) + c_2 r_2 (\mathbf{Gbest} - \mathbf{X}_{i}^{(t)})
     \label{eq7}
\end{equation}

\begin{equation}
    \mathbf{X}_{i}^{(t+1)} = \mathbf{X}_{i}^{(t)} + \mathbf{V}_{i}^{(t+1)}
     \label{eq8}
\end{equation}
where the learning factors are $c_1 = c_2 = 2.0$; $r_1, r_2 \sim \mathcal{U}(0, 1)$ are random perturbation matrices following a uniform distribution, which endow the algorithm with the randomness to continuously oscillate and avoid local extrema in the multidimensional space. The velocity and position are ultimately constrained within legitimate ranges through boundary constraint functions.

\subsection{The proposed GPSO}
To guarantee global exploration capability in the early search stages and accelerate local exploitation in the later stages, GPSO introduces a dynamic inertia weight $w(t)$ that decreases linearly with the iteration count $t$:
\begin{equation}
    w(t) = w_{max} - \frac{w_{max} - w_{min}}{Iter_{max}} \cdot t
     \label{eq9}
\end{equation}
where inertia weight is $w_{max} = 0.9$, and the minimum inertia weight is $w_{min} = 0.4$.

GPSO refines the initialization phase by introducing a Distance-based Heuristic strategy. The normalized distance vector $\mathbf{D}_{norm}$ from all target points to the takeoff point $H$ is calculated:
\begin{equation}
    \mathbf{D}_{norm} = \frac{\mathbf{D}_{home} - \min(\mathbf{D}_{home})}{\max(\mathbf{D}_{home}) - \min(\mathbf{D}_{home}) + \epsilon}
     \label{eq10}
\end{equation}

During initialization, GPSO forces half of the particles to adopt a heuristic distribution with distance priors, while retaining a uniform random perturbation of $0 \sim 0.3$ to maintain population diversity; the other half of the particles remain in a purely random state:
\begin{equation}
    \mathbf{X}_{i} = \mathbf{D}_{norm} + \mathcal{U}(0, 0.3), \quad \forall i \in \left[1, \frac{Pop}{2}\right)
     \label{eq11}
\end{equation}

\subsection{The proposed IPSO}
First, during the population initialization phase, a Tent Chaotic Map is introduced to completely replace pseudo-random numbers, ensuring that the initial particles possess a higher degree of ergodicity within the global search space. A minor perturbation of $\mathcal{U}(-0.001, 0.001)$ is superimposed to avoid fixed points:
\begin{equation}
    \mathbf{X}_{i,j} = \begin{cases} 2\mathbf{X}_{i-1,j}, & \mathbf{X}_{i-1,j} < 0.5 \\ 2(1 - \mathbf{X}_{i-1,j}), & \mathbf{X}_{i-1,j} \geq 0.5 \end{cases} + \mathcal{U}(-0.001, 0.001)
     \label{eq12}
\end{equation}

Second, after executing the conventional position update $\mathbf{X}_{i}^{(t+1)} = \mathbf{X}_{i}^{(t)} + \mathbf{V}_{i}^{(t+1)}$, IPSO innovatively concatenates a non-linear Lévy perturbation position update equation \cite{LSWOA} \cite{ASKSSA}. The Mantegna algorithm is utilized to generate a random step size $step$ following a Lévy distribution:
\begin{equation}
    step = \frac{u}{|v|^{1/\beta}}, \quad u \sim \mathcal{N}(0, \sigma^2), v \sim \mathcal{N}(0, 1)
     \label{eq13}
\end{equation}

To protect excellent path structures that have already converged, the system applies a linearly decaying step size weight $\alpha(t)$ and a random mask matrix $\mathbf{M}_{mask}$ following a $25\%$ probability, formulating the final secondary position perturbation equation:

\begin{equation}
    \alpha(t) = 0.05 \cdot (1 - \frac{t}{Iter_{max}})
     \label{eq14.5}
\end{equation}
\begin{equation}
    \mathbf{X}_{i}^{(t+1)} = \mathbf{X}_{i}^{(t+1)} + \alpha(t) \cdot step \cdot \mathbf{M}_{mask}
     \label{eq14}
\end{equation}

This mechanism ensures that while particles are engaged in local exploitation, they still retain the capability to execute long-range random jumps (Lévy Flight Mutation), thereby fundamentally suppressing the phenomenon of premature convergence.

\subsection{Optimal Selection Mechanism}
This framework innovatively introduces an algorithmic competition mechanism. The system records in parallel the convergence results of the original VLM sequence (Raw), PSO, GPSO, and IPSO. Through strict cost value comparison, the absolute optimal sequence with the shortest global distance is extracted:
\begin{equation}
    \mathcal{S}_{opt} = \arg\min_{\mathcal{S} \in \{Raw, PSO, GPSO, IPSO\}} F(\mathcal{S})
     \label{eq15}
\end{equation}

\subsection{Action Generation and MAVLink Deployment}
After obtaining the optimal GPS coordinate sequence, unlike traditional VLA systems that rely on network-side large language models to generate control code (which easily leads to high latency and action hallucinations), the ARIES-Mission2 system constructs a local deterministic mission parsing function for the \textit{\textbf{Mission Encapsulator}}. This module directly encapsulates the optimal physical coordinates with preset UAV flight parameters (e.g., cruising altitude of 100m, landing commands, etc.) to locally generate a standard MAVLink (QGC WPL 110) waypoint mission file within milliseconds. This purely localized deployment strategy not only guarantees extremely high computational efficiency but also achieves absolute physical safety in planning and control, allowing it to be directly uploaded to the UAV flight control system for execution.

\section{Experiments}
\begin{figure}[htbp]
    \centering
    \includegraphics[width=0.25\textwidth]{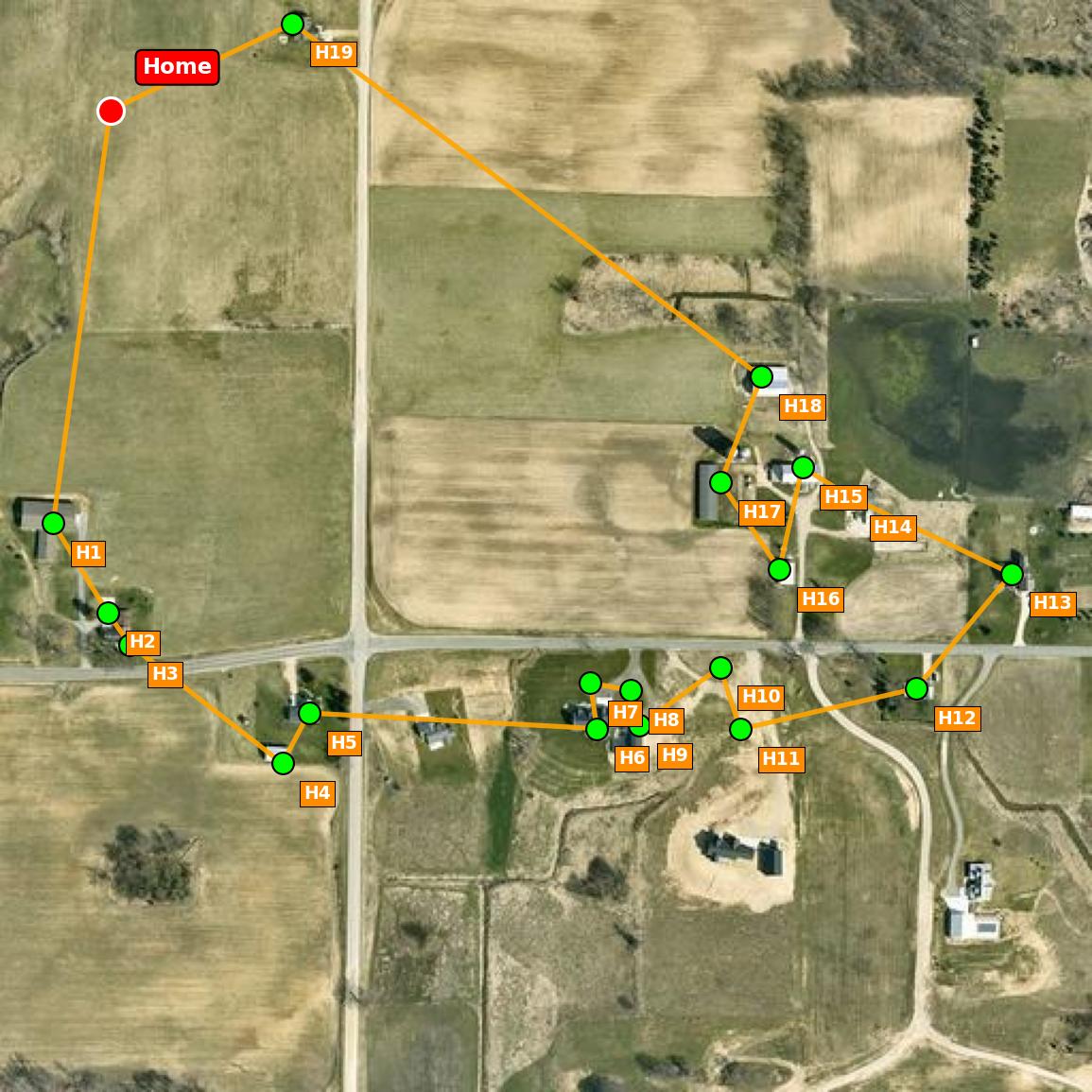}
    \caption{One of the flight trajectories manually planned by a human expert (HumanPlan).}
    \label{human_example}
\end{figure}

All experiments were conducted on a workstation equipped with an Intel Xeon E5-2698 v4 CPU (256\,GB RAM) and a single NVIDIA Tesla V100 GPU (32\,GB). To comprehensively evaluate the ARIES-Mission2 framework, we designed multi-dimensional comparative experiments. These benchmark the metaheuristic-integrated variants (PSO, GPSO, and IPSO) against both the pure vision-language-action baseline (ARIES-Mission2 w/o MAs) and manual planning by an experienced human expert (HumanPlan). Fig.~\ref{human_example} illustrates a manually planned flight trajectory.

To balance computational latency and global search capability, hyper-parameters for all metaheuristic algorithms within the TSP solver are uniformly configured. Specifically, the population size ($Pop$) is empirically set to 30, and the maximum number of iterations ($Iter_{max}$) is strictly defined as 150. These settings guarantee sufficient swarm diversity to escape local optima while achieving sub-second convergence, aligning perfectly with the real-time mission generation demands of the ARIES-Mission2 framework.

\subsection{Data Description}
\begin{figure}[htbp]
    \centering
    
    \begin{subfigure}{0.23\textwidth}
        \centering
        \includegraphics[width=\linewidth]{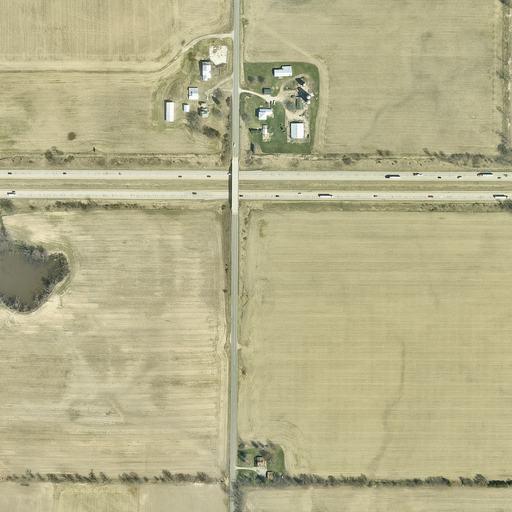}
        \caption{Task1}
        \label{task1}
    \end{subfigure}
    \hfill
    \begin{subfigure}{0.23\textwidth}
        \centering
        \includegraphics[width=\linewidth]{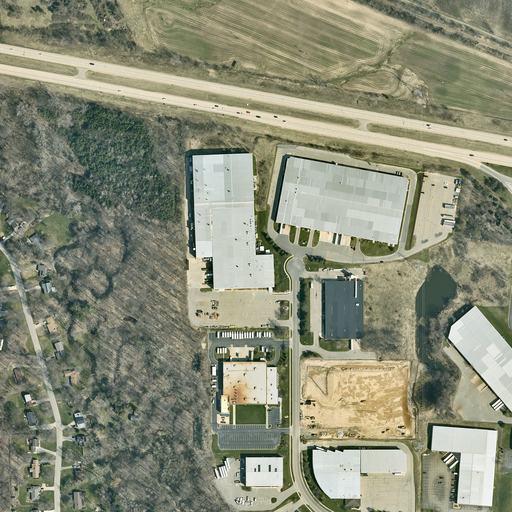}
        \caption{Task2}
        \label{task2}
    \end{subfigure}
    
    \caption{Task examples of high-resolution satellite imagery from the UAV-VLPA-nano-30 benchmark dataset.}
    \label{Examples}
\end{figure}

To validate the effectiveness of the ARIES-Mission2 framework, this study utilizes the UAV-VLPA-nano-30 benchmark dataset shown in Fig.~\ref{Examples} \cite{UAV-VLA}, which comprises 30 complex low-altitude flight tasks. Based on high-resolution real-world satellite imagery, this dataset encompasses diverse spatial topological structures of targets, including random distributions and dense clusters. Each task is paired with natural language instructions and precise geographic metadata (GPS boundaries), requiring the system to accurately extract target points and plan a closed-loop trajectory. The core evaluation metrics are the total flight distance ($Total$) and the average single-task flight distance ($Average$) required for the UAV to traverse all targets. These metrics objectively quantify the planning efficiency of the framework in realistic physical environments.

\subsection{Results Analysis}
\begin{figure}[htbp]
    \centering
    \includegraphics[width=0.45\textwidth]{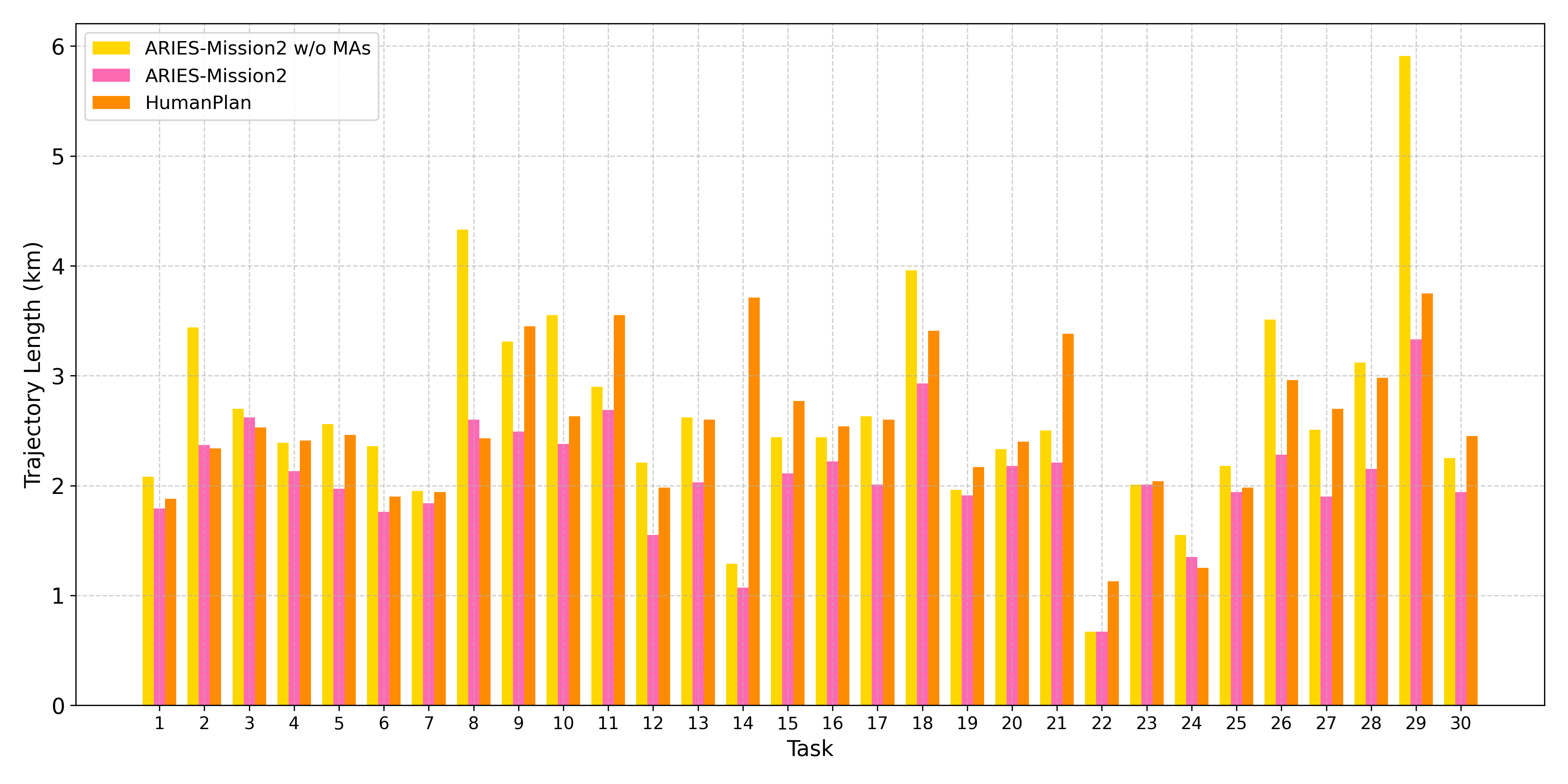}
    \caption{Quantitative comparison of trajectory lengths across 30 benchmark tasks. The chart contrasts the flight distances generated by the unoptimized pure VLA baseline (ARIES-Mission2 w/o MAs), the full ARIES-Mission2, and the manual expert baseline (HumanPlan).}
    \label{bar}
\end{figure}

\begin{figure}[htbp]
    \centering
    \includegraphics[width=0.25\textwidth]{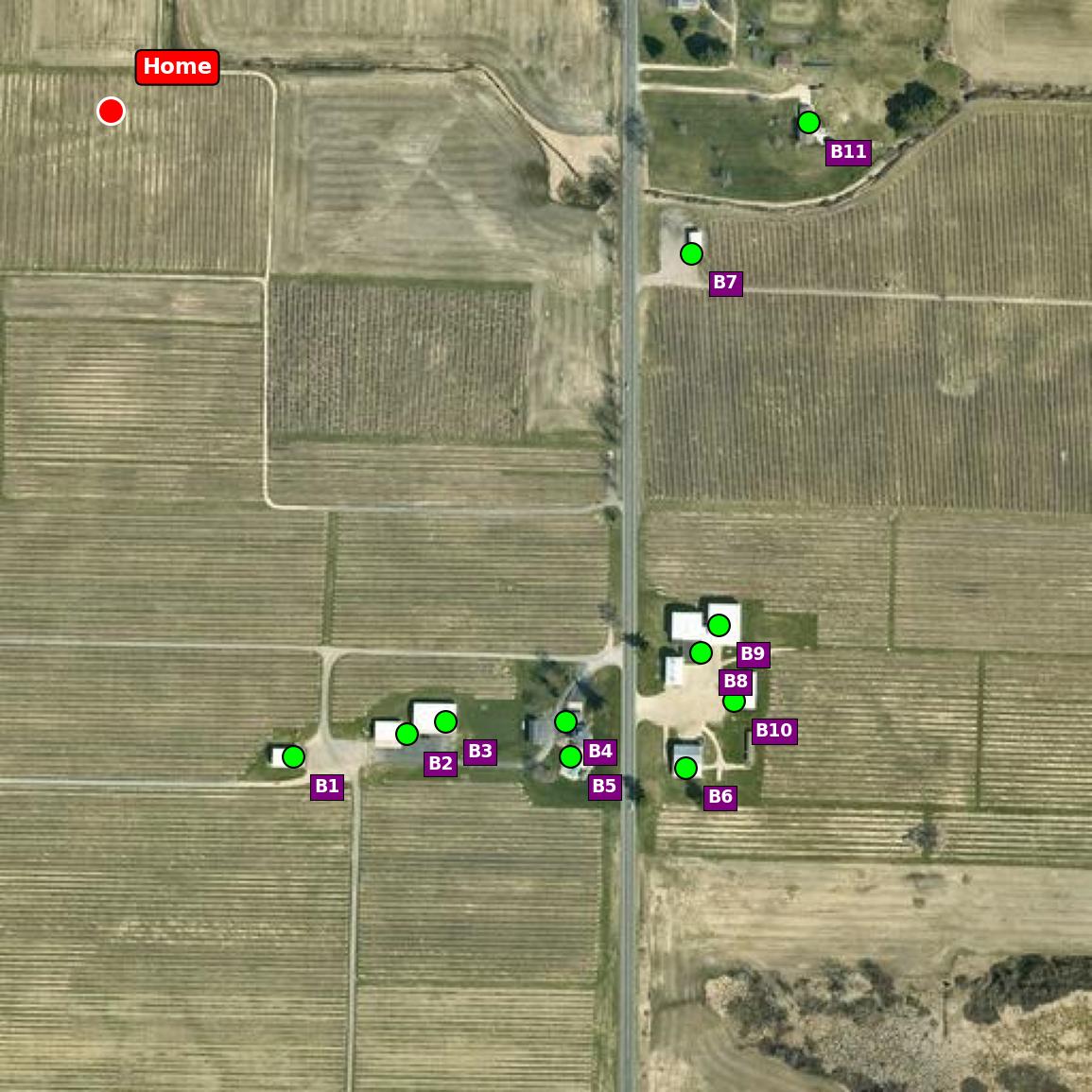}
    \caption{Target detection and spatial anchoring results for Task 6 using the DeepSeek-Molmo VLM module. All buildings are accurately identified within the satellite imagery under open-vocabulary conditions.}
    \label{building}
\end{figure}

\begin{figure}[htbp]
    \centering
    \includegraphics[width=0.25\textwidth]{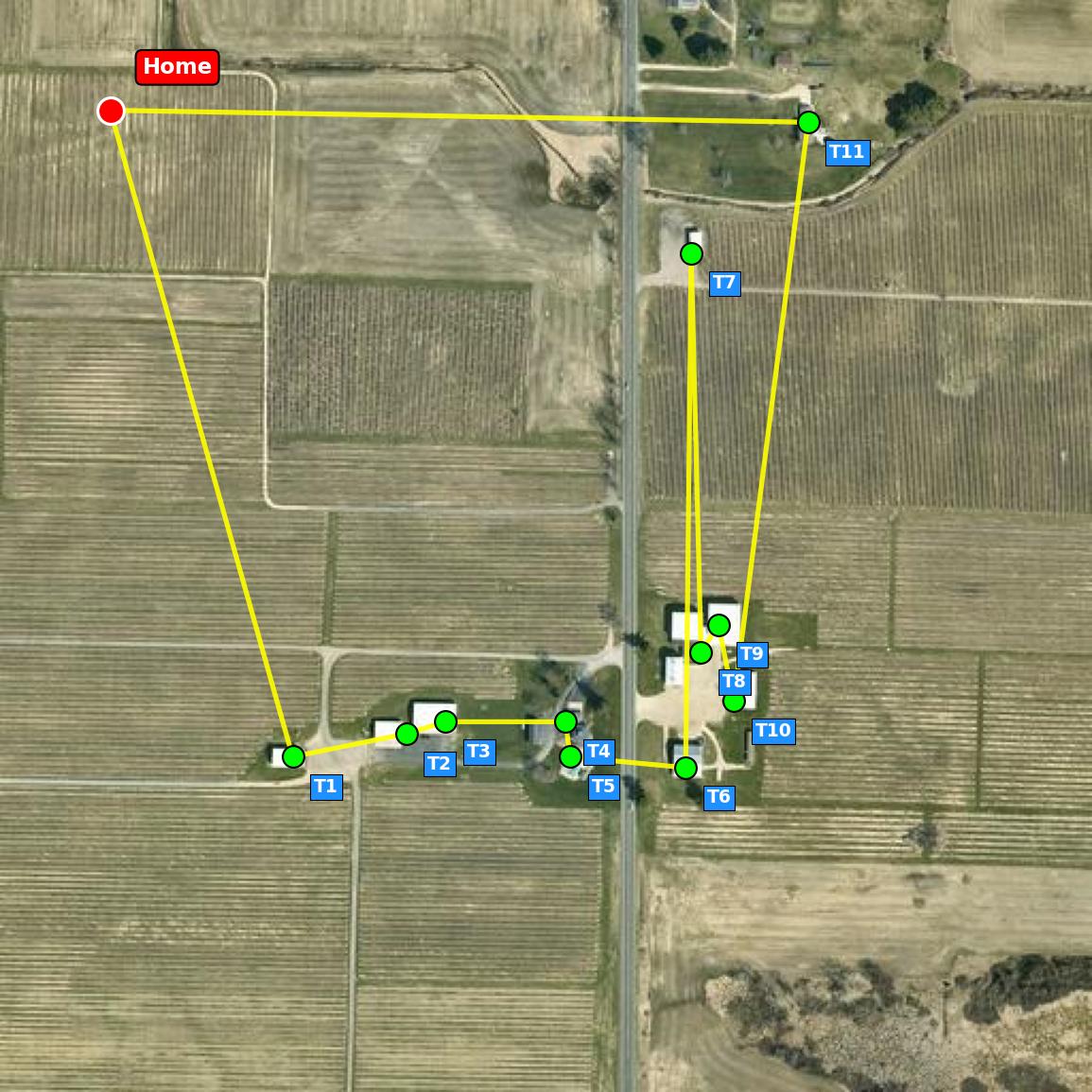}
    \caption{The chaotic and redundant flight trajectory generated by the pure VLA baseline (ARIES-Mission2 w/o MAs).}
    \label{raw}
\end{figure}

\begin{figure}[htbp]
    \centering

    \begin{subfigure}{0.48\columnwidth}
        \centering
        \includegraphics[width=\linewidth]{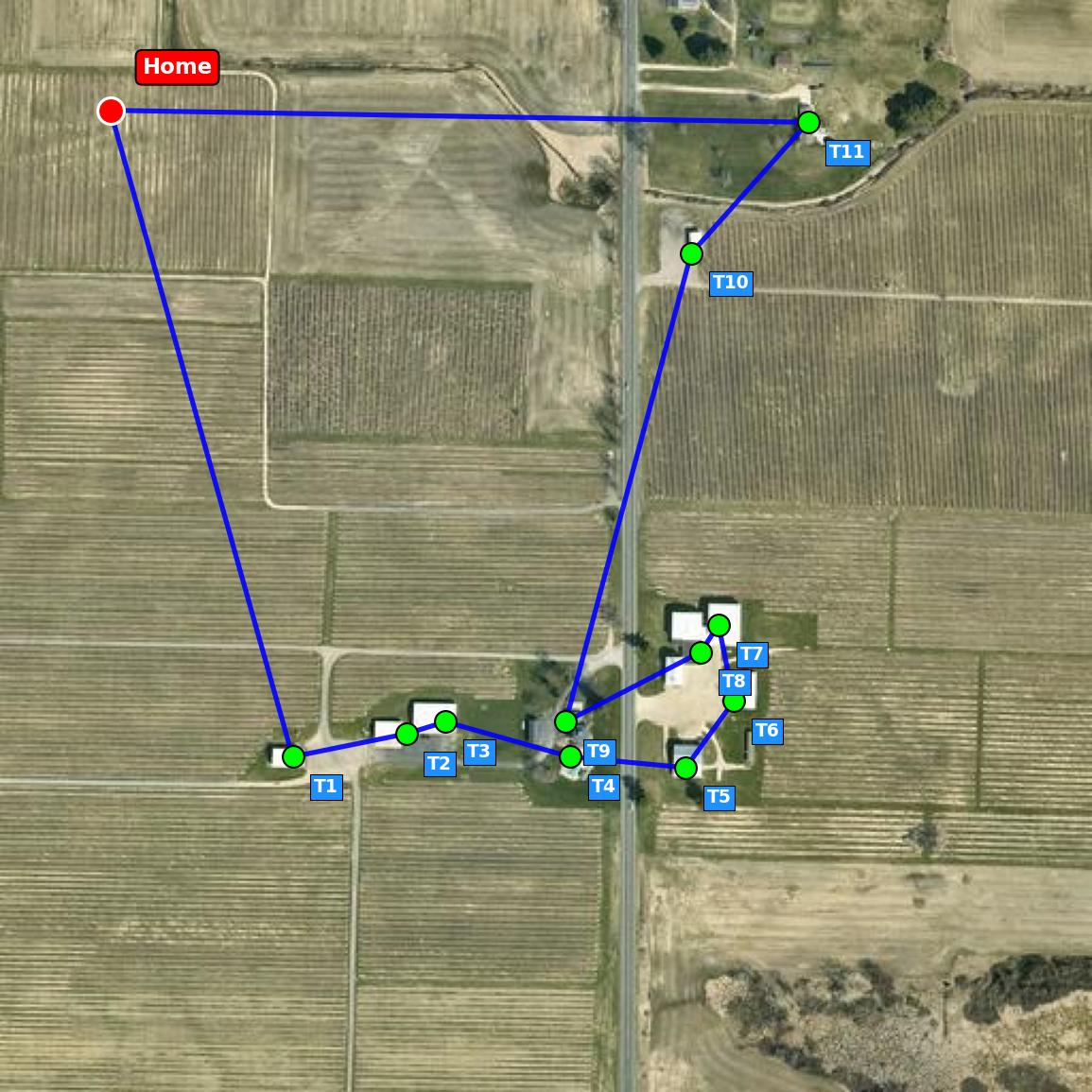}
        \caption{ARIES-Mission2-PSO}
        \label{pso}
    \end{subfigure}
    \hfill
    \begin{subfigure}{0.48\columnwidth}
        \centering
        \includegraphics[width=\linewidth]{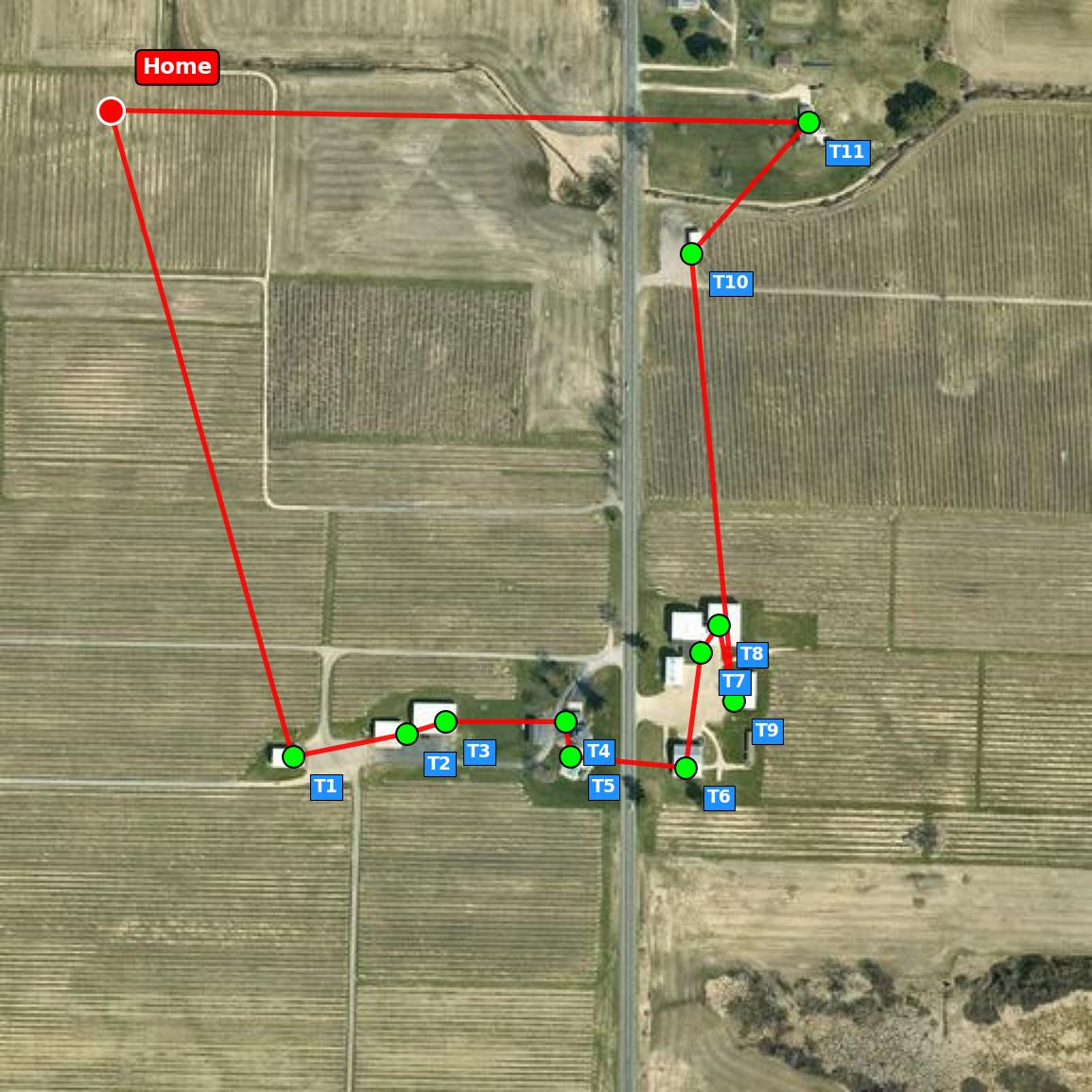}
        \caption{ARIES-Mission2-GPSO}
        \label{gpso}
    \end{subfigure}

    \vspace{0.6em}

    \begin{subfigure}{0.48\columnwidth}
        \centering
        \includegraphics[width=\linewidth]{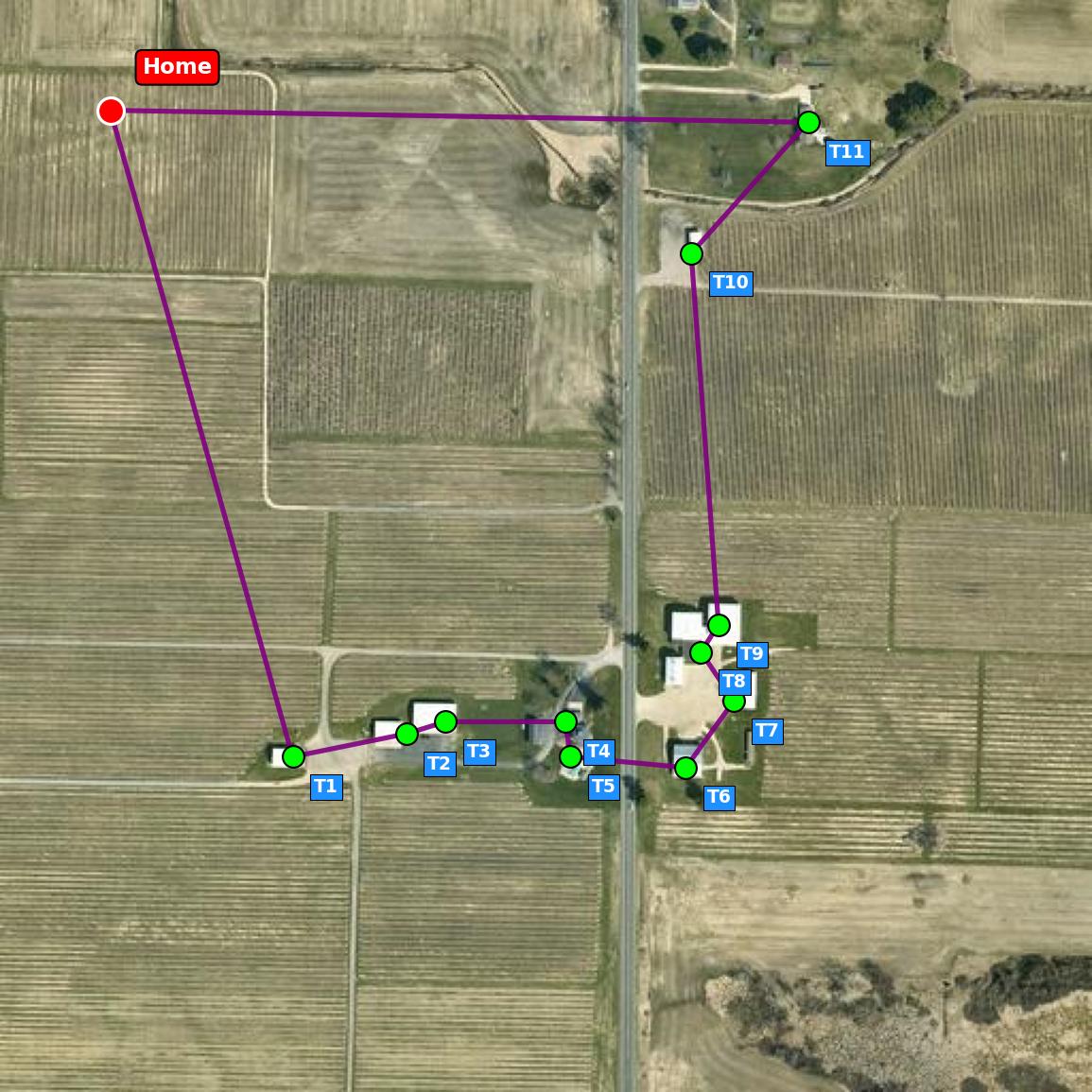}
        \caption{ARIES-Mission2-IPSO}
        \label{ipso}
    \end{subfigure}
    \hfill
    \begin{subfigure}{0.48\columnwidth}
        \centering
        \includegraphics[width=\linewidth]{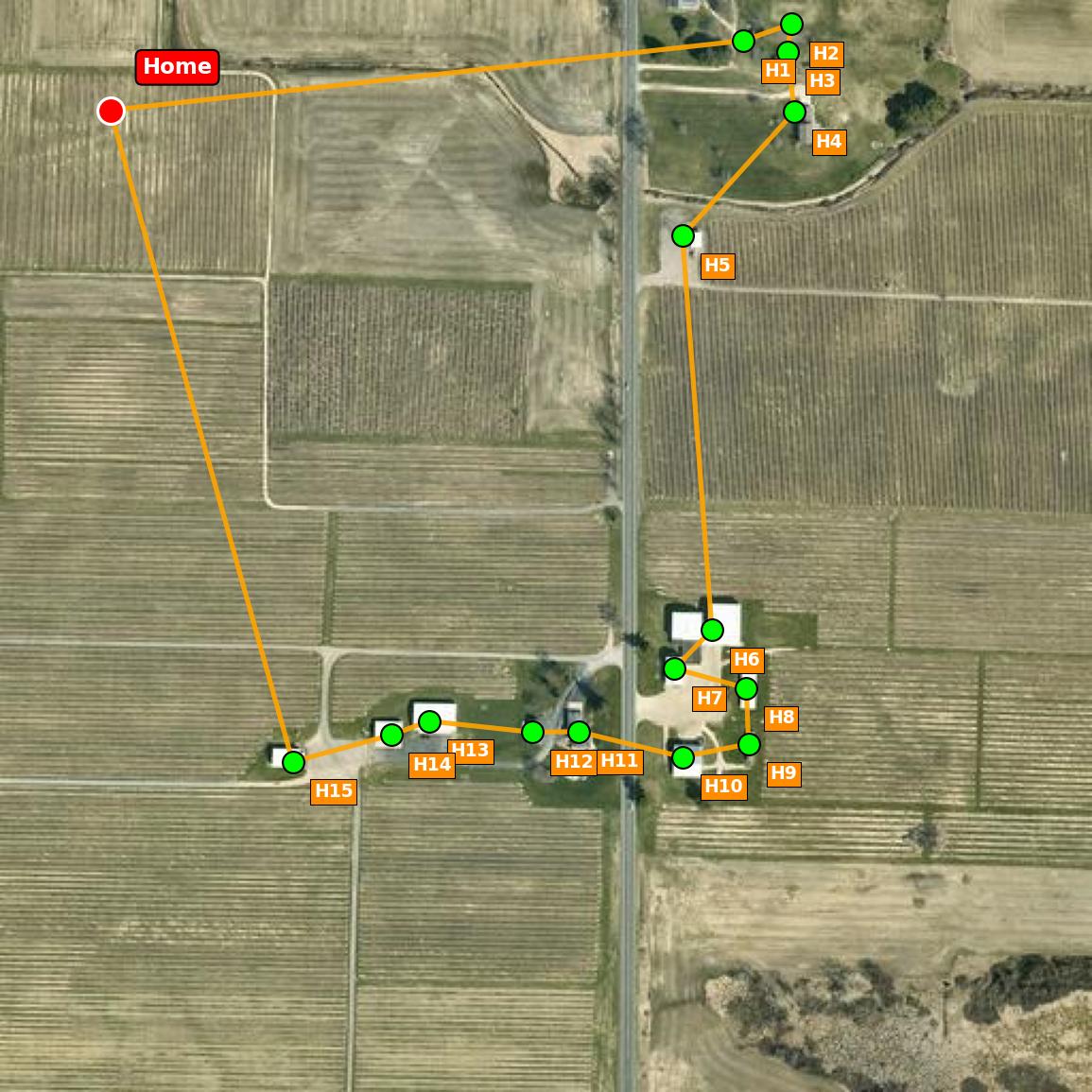}
        \caption{HumanPlan}
        \label{human}
    \end{subfigure}

    \caption{Optimized flight trajectories for Task 6 generated by different planning strategies. 
    PSO denotes the trajectory obtained using the standard PSO-based TSP solver, GPSO and IPSO denote enhanced PSO variants, while HumanPlan represents a baseline trajectory manually planned by an experienced human operator based on spatial intuition.}
    \label{fig:trajectory_comparison}
\end{figure}

The comparative results are shown in Table~\ref{tab:comparison} and Fig.~\ref{bar}. The pure VLA baseline without a TSP Solver (ARIES-Mission2 w/o MAs) yielded the poorest performance (79.66 km total distance). To illustrate, we take Task 6 as an example. Fig.~\ref{building} demonstrates the targets detected by the VLA model. Next, the VLA module sequentially connects the detected targets to form a closed-loop path, as shown in Fig.~\ref{raw}. This constitutes the tasks and trajectories generated by ARIES-Mission2 without MAs. The VLA baseline generates chaotic, backtracking trajectories, exposing MLLMs' severe deficiencies in optimal physical routing despite strong semantic perception. Conversely, human experts (HumanPlan) used spatial intuition to constrain the distance to 69.00 km, as illustrated in Fig.~\ref{human}. However, the full ARIES-Mission2 scheme (with the metaheuristic TSP Solver) drastically compressed the total distance to 62.43 km (2.08 km/task average). This reduces flight distance by 21.6\% compared to the VLA baseline and 9.5\% against human experts, demonstrating the framework's transcendence in complex spatial planning.

\begin{table}[htbp]
    \centering
    \caption{Comparison of Trajectory Lengths (km) across 30 Tasks. $Total$ stands for total distance of each method. $Average$ stands for average distance of each method.}
    \label{tab:comparison}
    \resizebox{\columnwidth}{!}{%
    \begin{tabular}{ccccccc}
        \toprule
        \textbf{Task} & \textbf{AM2-PSO} & \textbf{AM2-GPSO} & \textbf{AM2-IPSO} & \textbf{AM2 (Optimal)} & \textbf{AM2 w/o MAs} & \textbf{HumanPlan} \\
        \midrule
        1 & 1.81 & 1.79 & 1.81 & \textbf{1.79} & 2.08 & 1.84 \\
        2 & 2.44 & 2.37 & 2.48 & 2.37 & 3.44 & \textbf{2.08} \\
        3 & 2.67 & 2.62 & 2.66 & 2.62 & 2.70 & \textbf{2.49} \\
        4 & 2.33 & 2.32 & \textbf{2.13} & \textbf{2.13} & 2.39 & 2.17 \\
        5 & 1.97 & 2.31 & \textbf{1.97} & \textbf{1.97} & 2.56 & 2.13 \\
        6 & 1.94 & 1.85 & \textbf{1.76} & \textbf{1.76} & 2.36 & 1.86 \\
        7 & 1.84 & 1.84 & \textbf{1.84} & \textbf{1.84} & 1.95 & 1.94 \\
        8 & 3.34 & 3.35 & 2.60 & 2.60 & 4.33 & \textbf{2.31} \\
        9 & 2.62 & \textbf{2.49} & 2.52 & \textbf{2.49} & 3.31 & 2.71 \\
        10 & 2.38 & 2.44 & 2.52 & 2.38 & 3.55 & \textbf{2.31} \\
        11 & 2.78 & 2.78 & \textbf{2.69} & \textbf{2.69} & 2.90 & 3.16 \\
        12 & 1.68 & 1.61 & \textbf{1.55} & \textbf{1.55} & 2.21 & 1.87 \\
        13 & \textbf{2.03} & 2.24 & 2.14 & \textbf{2.03} & 2.62 & 2.30 \\
        14 & \textbf{1.07} & \textbf{1.07} & \textbf{1.07} & \textbf{1.07} & 1.29 & 3.71 \\
        15 & 2.14 & 2.21 & \textbf{2.11} & \textbf{2.11} & 2.44 & 2.39 \\
        16 & 2.28 & \textbf{2.22} & 2.26 & \textbf{2.22} & 2.44 & 2.54 \\
        17 & \textbf{2.01} & \textbf{2.01} & \textbf{2.01} & \textbf{2.01} & 2.63 & 2.49 \\
        18 & 3.59 & 2.96 & \textbf{2.93} & \textbf{2.93} & 3.96 & 3.01 \\
        19 & 1.91 & 1.91 & 1.91 & 1.91 & 1.96 & \textbf{1.77} \\
        20 & 2.19 & 2.19 & \textbf{2.18} & \textbf{2.18} & 2.33 & 2.40 \\
        21 & 2.25 & \textbf{2.21} & 2.24 & \textbf{2.21} & 2.50 & 2.63 \\
        22 & \textbf{0.67} & \textbf{0.67} & \textbf{0.67} & \textbf{0.67} & \textbf{0.67} & 1.13 \\
        23 & \textbf{2.01} & \textbf{2.01} & \textbf{2.01} & \textbf{2.01} & 2.01 & 2.04 \\
        24 & \textbf{1.35} & \textbf{1.35} & \textbf{1.35} & \textbf{1.35} & 1.55 & 1.25 \\
        25 & \textbf{1.94} & \textbf{1.94} & \textbf{1.94} & \textbf{1.94} & 2.18 & 1.98 \\
        26 & 2.40 & 2.67 & \textbf{2.28} & \textbf{2.28} & 3.51 & 2.33 \\
        27 & \textbf{1.90} & \textbf{1.90} & \textbf{1.90} & \textbf{1.90} & 2.51 & 2.12 \\
        28 & 2.19 & 2.24 & \textbf{2.15} & \textbf{2.15} & 3.12 & 2.42 \\
        29 & 4.27 & 3.45 & \textbf{3.33} & \textbf{3.33} & 5.91 & 3.50 \\
        30 & \textbf{1.94} & 2.15 & 2.19 & \textbf{1.94} & 2.25 & 2.12 \\
        \midrule
        \textbf{Total} & 65.95 & 65.17 & 63.20 & \textbf{62.43} & 79.66 & 69.00 \\
        \textbf{Average} & 2.20 & 2.17 & 2.11 & \textbf{2.08} & 2.66 & 2.30 \\
        \bottomrule
        \multicolumn{7}{l}{\footnotesize * AM2 stands for ARIES-Mission2.} \\
    \end{tabular}%
    }
\end{table}

\begin{figure}[htbp]
    \centering
    \includegraphics[width=0.35\textwidth]{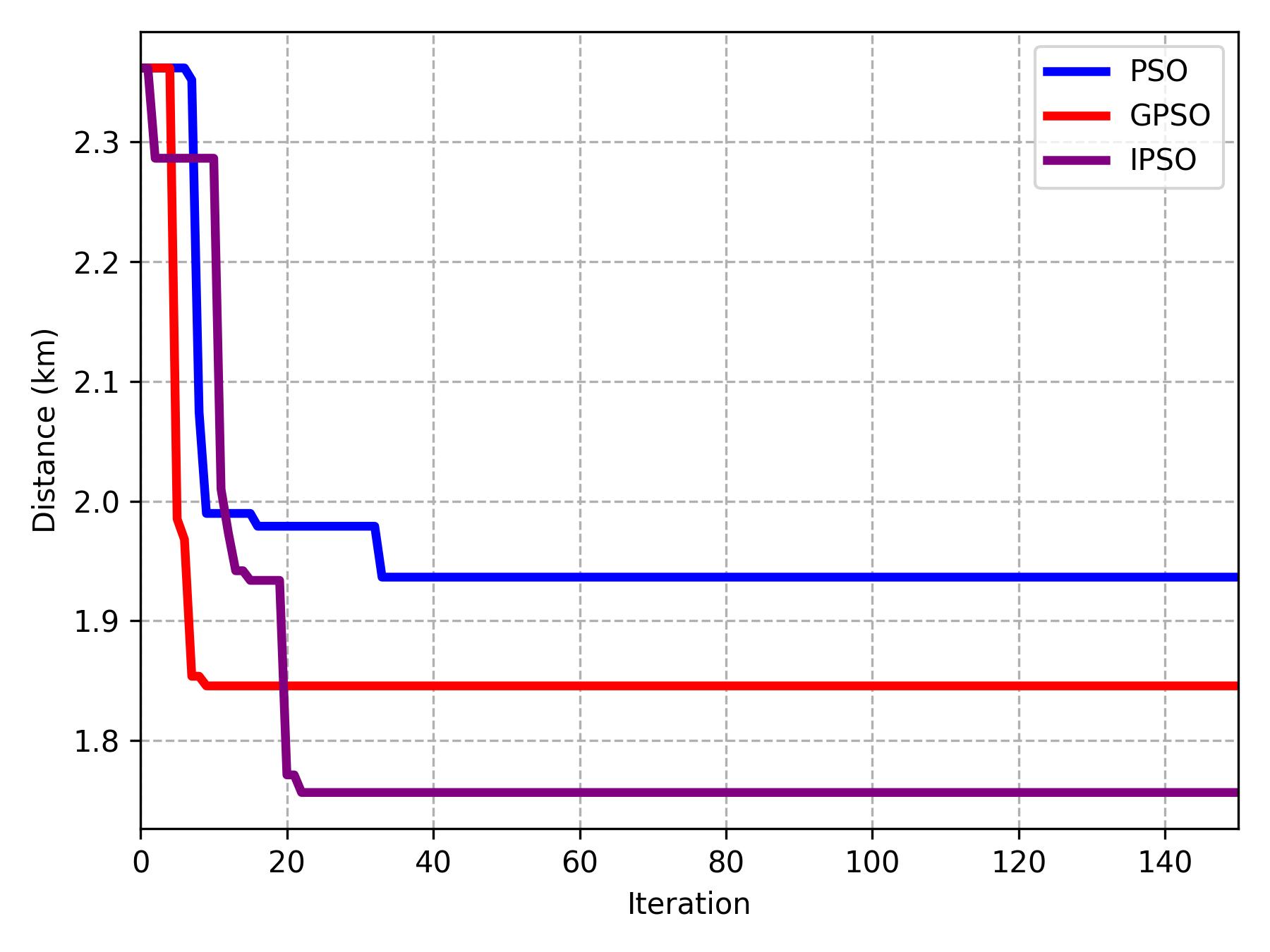}
    \caption{Comparative convergence curves of the metaheuristic algorithms (PSO, GPSO, and IPSO) in Task 6.}
    \label{iter}
\end{figure}

Algorithmic ablation using Task 6 (Fig.~\ref{iter}) elucidates the solvers' dynamical differences. Standard PSO (Fig.~\ref{pso}) shows weak global search due to random initialization, falling into a local optimum (1.94 km) behind the human expert (1.86 km). GPSO (Fig.~\ref{gpso}), using a distance heuristic, converges rapidly to surpass human planning at 1.85 km but quickly stagnates. In contrast, IPSO (Fig.~\ref{ipso}) demonstrates outstanding anti-premature convergence. By leveraging Levy Flight's non-linear mutation, IPSO successfully escapes local extrema traps, securing the global absolute optimal trajectory of 1.76 km. This validates IPSO's indispensability for high-dimensional flight tasks.

\begin{table}[htbp]
\centering
\caption{Comparison of computational latency between ARIES-Mission2 components and human expert planning across 30 tasks.}
\label{tab:time_comparison}
\renewcommand{\arraystretch}{1.2}
\resizebox{\linewidth}{!}{
\begin{tabular}{llcc}
\toprule
\textbf{Method} & \textbf{Pipeline Component} & \textbf{Total Time (s)} & \textbf{Avg. Time/Task (s)} \\
\midrule
Human Expert & Manual Route Planning & 2100.00 & 70.00 \\
\midrule
\multirow{3}{*}{ARIES-Mission2}
& VLM Visual Inference & 570.60 & 19.02 \\
& TSP Solver & 4.80 & 0.16 \\
\cmidrule{2-4}
& Complete Framework & \textbf{575.40} & \textbf{19.18} \\
\bottomrule
\end{tabular}
}
\end{table}

\begin{figure}[htbp]
    \centering
    \includegraphics[width=0.35\textwidth]{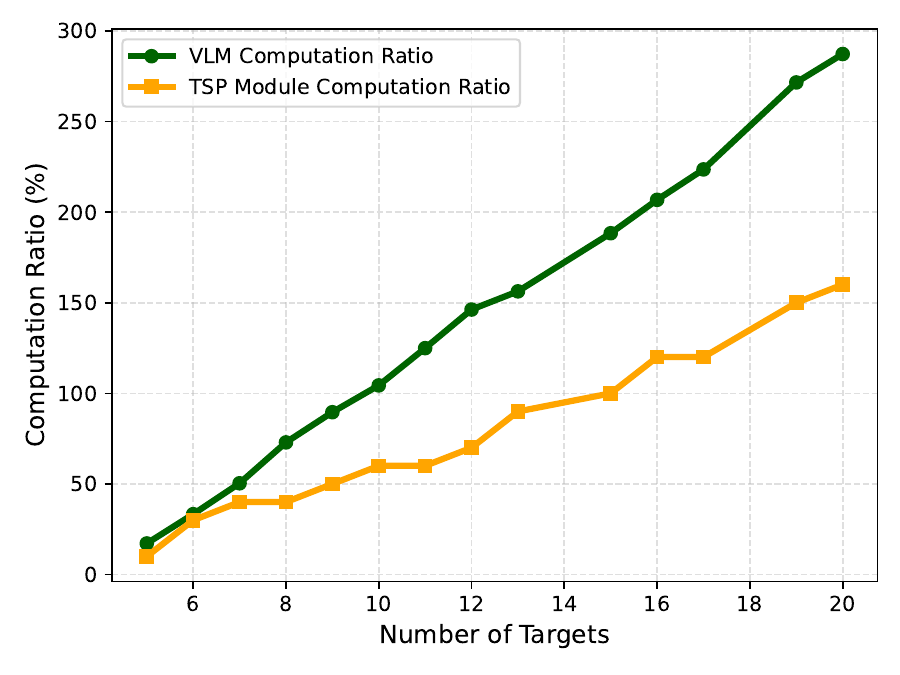}
    \caption{Computational scalability analysis showing the computation ratio relative to the number of targets ($N$).}
    \label{tbzz}
\end{figure}

To evaluate scalability, we analyzed the latency growth rate (Computation Ratio) of the VLM and TSP Solver modules as targets increase, using the minimum target count (4 points) as a baseline (Fig.~\ref{tbzz}). As targets scale to 20, VLM computation time shows steep linear growth, approaching a 300\% latency increase. In contrast, the TSP solver's overhead growth remains stable at around 160\%. This disparity highlights that the front-end VLM's visual inference is the core computational bottleneck for real-time performance. Conversely, the metaheuristic TSP solver maintains high efficiency and outstanding scalability, fully meeting the demands for large-scale, real-time mission planning in seconds.

\section{Conclusion}
Addressing the deficiencies of multimodal large models in spatial optimization capabilities during wide-area UAV missions, this paper proposes a novel VLA framework ARIES-Mission2 for fast aerial mission generation that decouples visual semantic perception from low-level physical optimization. The front end of the system utilizes a large model to accurately extract the true geographic coordinates of targets, while the back end innovatively integrates multiple metaheuristic algorithms to lock in the global optimal flight route through dynamic algorithmic competition, locally generating directly deployable MAVLink missions within seconds. Real-world benchmark tests demonstrate that with a total flight distance of 62.43 km, this system not only substantially rectifies the chaotic trajectory backtracking caused by the direct inference of the pure large model (79.66 km), but also significantly surpasses experienced human experts (69.00 km) in overall flight distance optimization. Furthermore, computational complexity analysis confirms that the back-end optimization solver exhibits exceptionally outstanding scalability, effectively circumventing the surging front-end computational bottleneck associated with an increasing number of targets. Overall, this paper provides a novel paradigm for the highly efficient real-world implementation of multimodal large models in complex physical spaces. Future work will further introduce 3D terrain \cite{geossa} \cite{aco1} \cite{cicdwoa} and dynamic obstacle avoidance mechanisms \cite{navrl} \cite{kio} to advance the fully autonomous application of embodied intelligence in low-altitude airspaces \cite{ego} \cite{saga} \cite{yopo} \cite{drrt}.

\section*{Acknowledgment}
This work (MPU submission code: fca.13ea.330d.3) was supported by the Macao Polytechnic University (RP/FCA-06/2026) and Macao Science and Technology Development Fund (FDCT-MOST: 0018/2025/AMJ).

\end{document}